\documentclass[12pt,preprint]{aastex}
\usepackage{amssymb}
\usepackage{amsmath}
\usepackage{graphicx}
\usepackage{subfigure}
\usepackage{wrapfig}
\usepackage[colorlinks,linkcolor=blue,anchorcolor=green,citecolor=blue]{hyperref}

\begin{document}

\def\func#1{\mathop{\rm #1}\nolimits}
\def\unit#1{\mathord{\thinspace\rm #1}}

\title{Reverse Shock Emission and Ionization Break Out Powered by
Post-merger Millisecond Magnetars}
\author{Ling-Jun Wang\altaffilmark{1,2}, Zi-Gao Dai\altaffilmark{1,2}, and
Yun-Wei Yu\altaffilmark{3}}

\begin{abstract}
There is accumulating evidence that at least a fraction of binary neutron
star mergers result in rapidly spinning magnetars, with subrelativistic
neutron-rich ejecta as massive as a small fraction of solar mass. The ejecta
could be heated continuously by the Poynting flux emanated from the central
magnetars. Such Poynting flux could become lepton-dominated so that a
reverse shock develops. It was demonstrated that such a picture is capable
of accounting for the optical transient PTF11agg \citep{wang13b}. In this
paper we investigate the X-ray and ultraviolet (UV) radiation as well as the
optical and radio radiation studied by \cite{wang13b}. UV emission is
particularly important because it has the right energy to ionize the hot
ejecta at times $t\lesssim 600\unit{s}$. It is thought that the ejecta of
binary neutron star mergers are a remarkably pure sample of $r$-process
material, about which our understanding is still incomplete. In this paper
we evaluate the possibility of observationally determining the bound-bound
and bound-free opacities of the $r$-process material by timing the X-ray,
UV, and optical radiation. It is found that these timings depend on the
opacities weakly and therefore only loose constraints on the opacities can
be obtained.
\end{abstract}



\affil{\altaffilmark{1}School of Astronomy and Space Science, Nanjing University, Nanjing,
China; dzg@nju.edu.cn}

\affil{\altaffilmark{2}Key laboratory of Modern Astronomy and Astrophysics (Nanjing
University), Ministry of Education, Nanjing 210093, China}

\affil{\altaffilmark{3}Institute of Astrophysics, Central China Normal University, Wuhan
430079, China; yuyw@mail.ccnu.edu.cn}

\section{Introduction}

\label{sec:intro}

Compact binary mergers are of great astrophysical importance as the primary
sources for the upcoming next generation ground-based gravitational wave
(GW) detectors \citep[see][for recent reviews]{abadie10,bartos13}. The
detection of GW can be confirmed if accompanying electromagnetic (EM)
signals are detected at the same time.

It has been believed that the mergers of compact binaries, i.e., double
neutron stars (NSs) or a NS with a stellar-mass black hole, are the
progenitors of short gamma-ray bursts %
\citep[SGRBs;][]{paczynski86,eichler89,barthelmy05,fox05,gehrels05,rezzolla11}%
. The evidence for such a scenario has recently gained strong support when
\cite{berger13} and \cite{tanvir13} discovered an $r$-process kilonova
associated with the SGRB 130603B\footnote{\cite{jin13} put forward an
alternative scenario based on a two-component jet model. The non-detection
of late radio emission of GRB 130603B \citep{fong14} does not favor such an
interpretation though. Despite of this fact, other possibilities %
\citep{fan13,takami14} cannot be ruled out on current observational ground.}%
. Kilonova, powered radioactively by the neutron-rich material ejected
during the coalescence of compact binaries, was first analytically predicted
by \cite{li98} and then intensely studied by many authors %
\citep{kulkarni05,rosswog05,metzger10b,roberts11,metzger12}. Other EM
signals include radio afterglows \citep{nakar11,metzger12,piran13,rosswog13}
and possible X-ray emission produced by the interaction of the NS
magnetospheres during the inspiral and merger process \citep{palenzuela13}.

In a popular scenario, the hypermassive remnant of binary neutron star (BNS)
merger collapses into a black hole on a timescale of $\sim 200\unit{ms}$\ %
\citep{rev-bin}. Models involving a black hole surrounded by a
hyperaccretion disc have been proposed %
\citep{Popham99,Narayan01,Kohri02,Xue13,Kawanaka13} as the central engine of
GRBs. Several authors \citep{dai06,zhang13}, however, suggest that a
highly-magnetized, rapidly spinning stable NS (magnetar) could form during
the coalescence of BNSs. The evidence for magnetar formation following some
SGRBs is accumulated during the past decade %
\citep{dai06,fan06,gao06,rowlinson10,rowlinson13,Dai12,wang13b,Gompertz14}.
In addition, magnetar activity could also be responsible for the statistical
properties of X-ray flares from some GRBs \citep{wang13a}. Recent
simulations in numerical relativity \citep[e.g.][]{magnetar-sim} confirmed
that formation of a stable long-lived magnetar is a plausible scenario.
However, in some cases the presence of such stable long-lived magnetars has
been ruled out by radio observations, as in the cases of GRB 050724 and GRB
060505 \citep{metzger-bower14}.

\cite{zhang13} recently proposed that if hypermassive millisecond magnetar
resulting from a coalescence of BNSs survives for sufficiently long time, a
significant fraction of NS-NS mergers would be discovered as bright X-ray
transients associated with GW bursts without SGRB association. \cite{gao13}
further studied the broadband EM signals of forward shock (FS) driven by the
ejecta based on the energy injection scenario \citep{dai98a,dai98b,Zhang01}.
Subsequently, \cite{yu13} studied the supernova-like EM signals of ejecta
(dubbed merger-nova) powered by the Poynting flux from post-merger magnetars
by assuming the complete absorption of Poynting flux by ejecta. \cite%
{wang13b}, on the other hand, analytically studied the reverse shock (RS)
emission powered by the central millisecond magnetars based on the fact that
the Poynting flux is more likely to become lepton-dominated
\citep[$e^+e^-$
pairs;][]{coroniti90,michel94,dai04,Yu07}. Such a scenario successfully
interprets observational data of the optical transient PTF11agg discovered
by the Palomar Transient Factory group \citep{cenko13,wang13b}\footnote{\cite%
{Wu14} proposed an alternative interpretation for this optical transient
within the framework of a magnetar-powered blast wave.}. More recently \cite%
{Metzger14} investigated the effects of pair-creation and annihilation on
the optical and X-ray emission from the remnant of BNSs mergers powered by
stable millisecond magnetars.

One of the major concerns in the theory of BNS mergers is the uncertainty in
the optical and nuclear properties of the ejected matter, which have started
to attract attention only recently %
\citep{goriely11,korobkin12,barnes13,bauswein13,grossman14,tanaka13}. In
particular, the opacity of the ejecta was usually set as $\kappa =0.2\unit{cm%
}^{2}\unit{g}^{-1}$ in the previous studies %
\citep[e.g.,][]{li98,rosswog05,roberts11,yu13}, which was recently found to
be $\kappa \sim 10\unit{cm}^{2}\unit{g}^{-1}$ %
\citep{barnes13,kasen13,tanaka13,grossman14}. As a result of much higher
values of opacity, much redder and dimmer kilonova transients are expected,
with their peak shifted to infrared, just as in the case of event associated
with SGRB 130603B \citep{berger13,tanvir13}.

The ejecta from a neutron star merger (NSM) are thought to be a pure sample
of $r$-process material. Apart from the above theoretical studies of the $r$%
-process material, it will be particularly valuable to observationally
constrain the opacity of the $r$-process material. In this paper, we
evaluate the possibility to constrain the bound-free and bound-bound
opacities, which are strongly dependent on the nuclear composition of the
ejecta. The physical picture of this scenario is illustrated in Figure \ref%
{fig:Schematic}. High opacity of the ejecta is caused by the presence of
elements with half-filled $f$-shells, such as lanthanides and actinides
produced in the $r$-process \citep{kasen13}. As a result of high
(bound-bound) opacity, optical emission from RS would be blocked during the
early expansion of the ejecta. Early X-ray emission from the RS, on the
other hand, would be blocked only by the electron scattering opacity and
unaffected by the bound-bound transitions. Consequently, we can determine
the launch time of the Poynting flux more accurately, more definitively by
timing X-ray emission. UV radiation from RS, however, is of great importance
because it has the right energy to ionize the hot ejecta. Through a
fortunate observation of X-ray, UV, and optical emission from RS powered by
a millisecond magnetar, we can determine the launch time of the central
magnetar (by X-ray observation), ionization break out by the UV radiation
and the time when the ejecta become optically transparent. The later two
times depend on the bound-free and bound-bound opacity, respectively.

In \cite{wang13b}, we demonstrated how our model consistently accounts for
the observed optical and radio properties of the PTF11agg transient. We
considered the case in which the spin-down time of the magnetar $T_{\mathrm{%
sd}}$\ exceeds the deceleration time of the blast wave $T_{\mathrm{dec}}$\ %
\citep[e.g., Case I of][]{gao13}. In Section \ref{sec:RS}, we further study
the other two cases. In our calculations, we adopt an improved version of
the blast wave dynamics, which is different from the one used in \cite{gao13}
and \cite{wang13b} in that it accounts for the different regions in the
blast wave. We compare different prescriptions for the dynamics in Section %
\ref{sec:RS}. In the same section we additionally investigate the effect of
the optical obscuration by the $r$-process material, neglected in \cite%
{wang13b}. We study the ionization effect of the UV radiation on the ejecta
in Section \ref{sec:Ion}. Finally, we discuss how to observationally
constrain the bound-bound and bound-free opacities in Section \ref%
{sec:discuss}. A summary is given in Section \ref{sec:conclusion}.

\section{Leptonized reverse shock emission}

\label{sec:RS}

Numerical simulations suggested that the ejecta from an NSM have a typical
velocity $v=0.1$-$0.3c$ and mass $M_{\mathrm{ej}}=10^{-4}$-$10^{-2}M_{\odot
} $ \citep{rezzolla10,hotokezaka13,rosswog13}. The resulting central compact
object, under the assumption that it can avoid collapse and persist for a
long time, would dissipate its rotational energy by launching a luminous
Poynting flux. Without the impact of the Poynting flux from the nascent
millisecond magnetar, the matter ejected during the merger would have
spatially extended morphology, which would soon have started to expand
homologously, despite the effects of radioactive heating \citep{rosswog14}.
With the onset of Poynting flux, the ejecta can be rapidly compressed into a
thin shell by the overpressure caused by the energy injection. The
overpressure will lead to a shock driving into the ejecta. The postshock
pressure is $p_{2}=2\gamma \rho _{1}v_{1}^{2}/\left( \gamma +1\right) $\ for
strong shock \citep{Landau87}, where $v_{1}$\ is the unshocked fluid
velocity relative to the contact discontinuity, $\rho _{1}$\ its density.
With $\rho _{1}=M_{\mathrm{ej}}\left( \frac{4}{3}\pi v^{3}t^{3}\right) ^{-1}$%
,\footnote{%
Here we ignore the initial size of the ejecta.} $p_{2}=L_{\mathrm{sd}}t\left[
3\times \frac{4}{3}\pi \left( v+v_{1}\right) ^{3}t^{3}\right] ^{-1}$\ and
the width of the initial ejecta $\Delta =v_{1}t$, we find the crossing time%
\begin{equation}
t_{\mathrm{cross}}=\frac{\Delta }{v_{1}}=\left( \frac{6\gamma }{\gamma +1}%
\frac{\Delta ^{2}M_{\mathrm{ej}}}{L_{\mathrm{sd}}}\right) ^{1/3}=0.09\unit{s}%
\Delta _{7}^{2/3}M_{\mathrm{ej},-4}^{1/3}L_{\mathrm{sd},47}^{-1/3}
\end{equation}%
and $v_{1}$\ at this time%
\begin{equation}
v_{1,\mathrm{cross}}=3.8\times 10^{-3}c\Delta _{7}^{1/3}M_{\mathrm{ej}%
,-4}^{-1/3}L_{\mathrm{sd},47}^{1/3}\ll v\approx 0.2c,
\end{equation}%
where $L_{\mathrm{sd}}$ is the spin-down luminosity of the magnetar. Here
the usual convention $Q=10^{n}Q_{n}$ is adopted. This crossing time is
negligible compared to the activity duration of the central magnetars,
justifying the model we are considering.

As mentioned in Section \ref{sec:intro}, the Poynting flux from millisecond
magnetars is more likely to become lepton-dominated. In developing a model
to account for the transient source PTF11agg, \cite{wang13b} adopted a
dynamics%
\begin{equation}
L_{0}\min \left( t,T_{\mathrm{sd}}\right) =\left( \gamma -\gamma _{\mathrm{ej%
},0}\right) M_{\mathrm{ej}}c^{2}+2\left( \gamma ^{2}-1\right) M_{\mathrm{sw}%
}c^{2},  \label{eq:PTF11agg-dyn}
\end{equation}%
which is different by a factor of 2 on the second term from that used by
\cite{gao13}. In the above equation $L_{0}=\xi L_{\mathrm{sd}}$ with $\xi $
the fraction of the Poynting flux catched by the ejecta, $\gamma $ the
Lorentz factor of the ejecta with initial Lorentz factor $\gamma _{\mathrm{ej%
},0}$, $M_{\mathrm{sw}}$ the mass swept up by the shock.

In this paper we would like to adopt an alternate dynamics similar to that
used by \cite{yu13} and to evaluate any differences between these two kinds
of dynamics. In this scenario, the ejecta gain energy because of the work
done by RS and likewise the FS gains energy by the work done by the ejecta.
As a result, the ejecta are sandwiched between RS and FS. The treatment of
RS is otherwise similar to that in \cite{wang13b}.

\begin{figure}[tbph]
\centering\includegraphics[width=0.95\textwidth,angle=0]{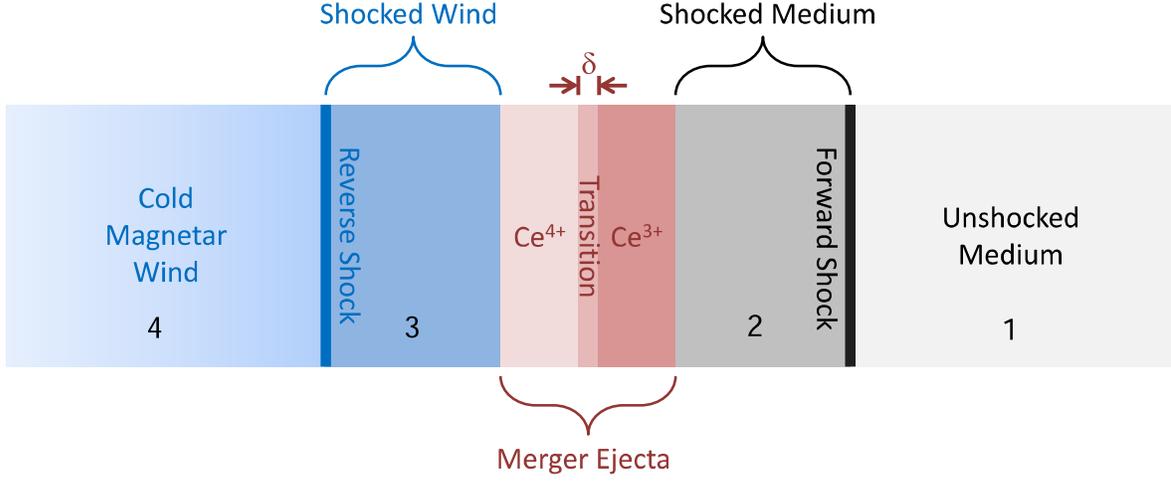}
\caption{Schematic illustration of the physical picture described in this
paper. The merger ejecta are catched up by the relativistic $e^{\pm }$ wind,
driving the surrounding medium to form forward shock. The magnetar wind, on
the other hand, is heated by the reverse shock. The ultraviolet radiation
from the reverse shock ionizes the thermally ionized Ce$^{3+}$ to Ce$^{4+}$.
$\protect\delta $ is the transition width between Ce$^{4+}$ zone and Ce$%
^{3+} $ zone. When the ejecta temperature drops below $T_{\mathrm{ion}}$,
viz. the temperature to thermally ionize Ce to Ce$^{3+}$, Ce$^{3+}$ begins
to recombine. Consequently, the ionization stage can only last for a limited
time. The Arabic numbers indicate different regions: region 1 is the
unshocked medium, region 2 is the shocked medium, region 3 is the shocked
wind, region 4 is the unshocked cold wind.}
\label{fig:Schematic}
\end{figure}

The total non-rest energy of the ejecta and shocked media (including FS and
RS) can be expressed as%
\begin{equation}
E_{k}=\gamma E_{3}^{\prime }+\left( \gamma -\gamma _{\mathrm{ej},0}\right)
M_{\mathrm{ej}}c^{2}+\gamma E_{\mathrm{ej,int}}^{\prime }+\left( {\gamma }%
^{2}-1\right) M_{\mathrm{sw}}c^{2},
\end{equation}%
where $E_{3}^{\prime }$ and $E_{\mathrm{ej,int}}^{\prime }$ are the
respective energies of the reverse-shocked wind (region 3) and ejecta in the
comoving frame.\ For the definition of regions 1-4, see Figure \ref%
{fig:Schematic} \citep[see also][]{dai04}. Here we neglect the rest energy
of region 3 because it is lepton-dominated and therefore very hot. In a way
similar to the generic dynamic model for GRB afterglow \citep{huang99}, the
dynamics can be derived as \citep[cf.][]{yu13}%
\begin{equation}
\frac{d\gamma }{dt}=\frac{\xi L_{\mathrm{sd}}+L_{\mathrm{rd}}-L_{\mathrm{ej,}%
e}-\gamma \mathcal{D}\left( \frac{dE_{3}^{\prime }}{dt^{\prime }}+\frac{dE_{%
\mathrm{ej,int}}^{\prime }}{dt^{\prime }}\right) -\left( {\gamma }%
^{2}-1\right) {c}^{2}\left( \frac{dM_{\mathrm{sw}}}{dt}\right) }{%
E_{3}^{\prime }+M_{\mathrm{ej}}c^{2}+E_{\mathrm{ej,int}}^{\prime }+2\gamma
M_{\mathrm{sw}}c^{2}},  \label{eq:dynamics-RS-work}
\end{equation}%
where $\mathcal{D}$ is the Doppler factor, $L_{\mathrm{rd}}$ the radioactive
luminosity of the ejecta in the observer frame, $L_{\mathrm{ej,}e}$ the
energy loss rate in ejecta due to thermal radiation. The evolution of
energies in region 3 and ejecta can be expressed as
\begin{eqnarray}
\frac{dE_{3}^{\prime }}{dt^{\prime }} &=&\xi L_{\mathrm{sd}}^{\prime }-p_{3}%
\frac{dV_{3,\mathrm{enc}}^{\prime }}{dt^{\prime }},  \label{eq:E3_derivative}
\\
\frac{dE_{\mathrm{ej,int}}^{\prime }}{dt^{\prime }} &=&p_{3}\frac{dV_{3,%
\mathrm{enc}}^{\prime }}{dt^{\prime }}-p_{\mathrm{ej}}\frac{dV_{\mathrm{ej},%
\mathrm{enc}}^{\prime }}{dt^{\prime }}+L_{\mathrm{rd}}^{\prime }-L_{\mathrm{%
ej},e}^{\prime },  \label{eq:E_ej_derivative}
\end{eqnarray}%
where $t^{\prime }$ is the time measured in the comoving frame, $L_{\mathrm{%
ej},e}^{\prime }$ the thermal radiation of ejecta. Here we have to discern
different volumes involved in this problem. See Figure \ref{fig:Schematic}
for reference. $V_{3,\mathrm{enc}}^{\prime }$\ and $V_{\mathrm{ej},\mathrm{%
enc}}^{\prime }$\ are the comoving volumes enclosed by region 3 and ejecta,
respectively. $V_{3}^{\prime }$\ and $V_{\mathrm{ej}}^{\prime }$, on the
other hand, are the volumes actually occupied by these two regions. In other
words, the volumes with the subscript $enc$\ denote the spherical volumes,
the volumes without such subscript denote the shell volumes. The pressures
in region 3 and ejecta in the comoving frame are $p_{3}=e_{3}/3$ and $p_{\mathrm{%
ej}}=E_{\mathrm{ej,int}}^{\prime }/3V_{\mathrm{ej}}^{\prime }$,
respectively. The width of the sandwiched ejecta can be determined by
assuming pressure balance $p_{3}=p_{\mathrm{ej}}$. Because the ejecta are
sandwiched in a thin shell, i.e., the width of ejecta (in the comoving
frame) $\Delta _{\mathrm{ej}}^{\prime }\ll r$, so that its volume $V_{%
\mathrm{ej}}^{\prime }\approx 0$, Equation $\left( \ref{eq:E_ej_derivative}%
\right) $ can be approximately written as%
\begin{equation}
\frac{dE_{\mathrm{ej,int}}^{\prime }}{dt^{\prime }}=L_{\mathrm{rd}}^{\prime
}-L_{\mathrm{ej},e}^{\prime }.
\end{equation}%
The comoving volume can be found as \citep{yu13}
\begin{equation}
\frac{dV_{3,\mathrm{enc}}^{\prime }}{dt^{\prime }}=4\pi r^{2}\xi \beta c,
\end{equation}%
with%
\begin{equation}
\frac{dr}{dt}=\frac{\beta c}{1-\beta }.
\end{equation}%
The radioactive luminosity in the comoving frame $L_{\mathrm{rd}}^{\prime
}=L_{\mathrm{rd}}/\mathcal{D}^{2}$ is determined according to Equation $%
\left( 4\right) $ of \cite{korobkin12}. The thermal energy $L_{\mathrm{ej}%
,e}^{\prime }$ emitted by the sandwiched ejecta in the comoving frame is
still expressed by Equation (7) in \cite{yu13}, while the luminosity light
curves at a particular observational frequency are modified as%
\begin{equation}
\nu L_{\nu }=\frac{1}{\max \left( \tau ,1\right) }\frac{8\pi ^{2}\mathcal{D}%
^{2}r^{2}}{h^{3}c^{2}}\frac{\Delta _{\mathrm{ej}}^{\prime }}{r}\frac{\left(
h\nu /\mathcal{D}\right) ^{4}}{\exp \left( h\nu /\mathcal{D}kT^{\prime
}\right) -1},
\end{equation}%
i.e., there is an extra factor $\Delta _{\mathrm{ej}}^{\prime }/r$ compared
with the solid spherical geometry.

Inspection of Equation $\left( \ref{eq:PTF11agg-dyn}\right) $\ indicates
that initially $\left( \gamma -\gamma _{\mathrm{ej},0}\right) M_{\mathrm{ej}%
}c^{2}\gg 2\left( \gamma ^{2}-1\right) M_{\mathrm{sw}}c^{2}$, the ejecta
will be accelerated linearly in time until $t=\min \left( T_{\mathrm{sd}},T_{%
\mathrm{dec}}\right) $, where the deceleration timescale $T_{\mathrm{dec}}$\
is determined by the condition $\left( \gamma -\gamma _{\mathrm{ej}%
,0}\right) M_{\mathrm{ej}}c^{2}=2\left( \gamma ^{2}-1\right) M_{\mathrm{sw}%
}c^{2}$\ \citep[see also][]{gao13}. By setting the spin-down timescale $T_{%
\mathrm{sd}}\sim T_{\mathrm{dec}}$, we will arrive at a critical ejecta mass %
\citep{gao13}%
\begin{equation}
M_{\mathrm{ej},c}\sim 10^{-3}M_{\odot
}n^{1/8}I_{45}^{5/4}L_{0,47}^{-3/8}P_{0,-3}^{-5/2}\xi ^{5/4},
\end{equation}%
where $I$\ is the moment of inertia of the neutron star. It is this critical
mass that defines the three cases considered by \cite{gao13}, i.e., Case I
for $T_{\mathrm{sd}}>T_{\mathrm{dec}}$, Case II for $T_{\mathrm{sd}}=T_{%
\mathrm{dec}}$, and Case III for $T_{\mathrm{sd}}<T_{\mathrm{dec}}$. Here we
show the corresponding results including RS in Figures \ref{fig:Case-I}-\ref%
{fig:Case-III}, where the spectral index of electrons in FS is set as $%
p^{FS}=2.3$, and the luminosity distance of the source is set as $%
D_{L}=10^{27}\unit{cm}$. The optical opacity is set as $\kappa =10\unit{cm}%
^{2}\unit{g}^{-1}$. For more discussions about the dependence of opacity on
wavelength, see Section \ref{sec:Ion}. To ease comparison between different
choices of dynamics, we selectively show in Figures \ref{fig:Case-I}a and %
\ref{fig:Case-I}b as dashed lines the results with the same parameters as
the solid lines but with dynamics expressed by Equation $\left( \ref%
{eq:PTF11agg-dyn}\right) $.

In the above calculations we set the Lorentz factor of the unshocked
electrons/positrons (region 4) as $\gamma _{4}=10^{4}$, as determined in the
literature \citep{atoyan99,dai04,wang13b}. The spectral index of $e^{+}e^{-}$
in RS is set as $p^{RS}=2.2$ \citep[see, e.g.,][]{wang13b}. There is an
additional complication concerning RS that should be mentioned. After the
shut-off of the central magnetar at $T_{\mathrm{sd}}$, the reverse shock
crosses region 4 so that there are no more $e^{+}e^{-}$ to be shocked.
Consequently, the electrons already cooled to a Lorentz factor $\gamma _{c}$
cannot be shocked again to a higher Lorentz factor. As a result, in the
analytical and numerical calculations, if we find the cooling Lorentz factor
increases in a time period after $T_{\mathrm{sd}}$, which is the case when $%
t>T_{N2}$ for Case I, we will set its value the same as before.

We find that the characteristic frequencies of synchrotron radiation of FS
are quite similar to that calculated by \cite{gao13} so that we do not show
them in the figures. Because the RS emission will be absorbed by the ejecta,
we show the RS emission with/without absorption in Panels (d) and (c)
respectively. It is in principle possible that a FS and a RS develop in the
ejecta upon the interaction with region 3 and region 2 respectively. In
practice, however, since the ejecta are very thin, the FS and RS in it can
last only for a transient while. Consequently we do not consider them here.

A first glimpse of Figure \ref{fig:Case-I}a shows that the evolution of
Lorentz factor with different choices of dynamics are very similar. However,
there are also differences that result in appreciable modification to the
evolution of characteristic frequencies (comparing the solid lines and
dashed lines in Figure \ref{fig:Case-I}b) and light curves, which are not
shown in Figure \ref{fig:Case-I}. The reason is that the ejecta absorb some
amount of energy, resulting in a heavier ejecta and therefore a delay of the
deceleration time $T_{\mathrm{dec}}$ by a factor of $2.1$. This further
results in a drop of $\nu _{m}$ (see Figure \ref{fig:Case-I}b), which is
otherwise identical to the situation where the ejecta do not absorb energy
(comparing the solid lines and dashed lines in Figure \ref{fig:Case-I}b).
The delay of $T_{\mathrm{dec}}$ also strengthens the FS since more ambient
media are swept up at $T_{\mathrm{dec}}$. Figure \ref{fig:Case-III}a shows
the coasting of Lorentz factor, but not so clear-cut as in Figure 4a of \cite%
{gao13}.

Although the energy absorption by ejecta has an appreciable effect on the
dynamics and therefore the light curves, it is not so significant as for the
case of merger-nova discussed by \cite{yu13}. We therefore conclude that a
simple model based on Equation $\left( \ref{eq:PTF11agg-dyn}\right) $\ and
used in \cite{wang13b}, properly reproduces qualitative features of the more
sophisticated one employed here, and therefore the use of this model in \cite%
{wang13b} is justified.

For completeness and also for the ease of future quantitative analysis, we
present the analytical results of RS based on the dynamics $\left( \ref%
{eq:PTF11agg-dyn}\right) $ for Case II and Case III below and their temporal
scaling indices in Table \ref{tbl:indices}
\citep[the corresponding results for Case
I can be found in][]{wang13b}. For Case II, the various time scales and the
peak Lorentz factor are%
\begin{eqnarray}
T_{\mathrm{N1}} &=&2.1\times 10^{-4}\unit{days}L_{0,49}^{-1}M_{\mathrm{ej}%
,-4} \\
T_{\mathrm{ct}} &=&2.8\times 10^{-3}\unit{days}L_{0,49}^{-2/3}M_{\mathrm{ej}%
,-4}^{5/6}\epsilon _{B,-1}^{1/6} \\
T_{ac} &\approx &T_{mc}=5.0\times 10^{-3}\unit{days}L_{0,49}^{-5/7}M_{%
\mathrm{ej},-4}^{6/7}\epsilon _{B,-1}^{1/7}\epsilon _{e}^{1/7}\gamma
_{4,4}^{1/7} \\
T_{\mathrm{dec}} &\simeq &T_{\mathrm{sd}} \\
T_{\mathrm{N2}} &=&2.4\times 10^{2}\unit{days}L_{0,49}^{1/3}T_{\mathrm{sd}%
,3}^{1/3}n^{-1/3} \\
T_{am2} &=&9.4\times 10^{2}\unit{days}L_{0,49}^{49/2}T_{\mathrm{sd}%
,3}^{49/2}M_{\mathrm{ej},-4}^{-45/2}\gamma _{4,4}^{-30}n^{1/2}\epsilon
_{B,-1}^{-5/2}\epsilon _{e}^{-25}  \notag \\
&&\times \left[ \left( p-1\right) \Gamma \left( \frac{3p+22}{12}\right)
\Gamma \left( \frac{3p+2}{12}\right) \right] ^{5}\left( \frac{p-1}{p-2}%
\right) ^{25} \\
\gamma _{\mathrm{sd}} &=&28L_{0,49}T_{\mathrm{sd},3}M_{\mathrm{ej}%
,-4}^{-1}+1,
\end{eqnarray}%
where $T_{am2}$ is the time when $\nu _{a}$ crosses $\nu _{m}$ at the second
time, which is very sensitive to many parameters because these two
frequencies almost have the same temporal evolution indices during the time
period $T_{N2}<t<T_{am2}$ (Table \ref{tbl:indices}, see also Figure \ref%
{fig:Case-II}b). The characteristic frequencies and the observed peak flux
are%
\begin{eqnarray}
\nu _{a,\mathrm{sd}} &=&1.7\times 10^{12}\unit{Hz}L_{0,49}^{-\left(
7p+10\right) /2\left( p+4\right) }T_{\mathrm{sd},3}^{-\left( 5p+12\right)
/\left( p+4\right) }  \notag \\
&&\times M_{\mathrm{ej},-4}^{4\left( p+2\right) /\left( p+4\right) }\epsilon
_{e}^{2\left( p-1\right) /\left( p+4\right) }\epsilon _{B,-1}^{\left(
p+2\right) /2\left( p+4\right) }\gamma _{4,4}^{2\left( p-2\right) /\left(
p+4\right) } \\
\nu _{m,\mathrm{sd}} &=&1.2\times 10^{12}\unit{Hz}L_{0,49}^{-7/2}\epsilon
_{B,-1}^{1/2}\epsilon _{e}^{2}\gamma _{4,4}^{2}\left( \frac{p-2}{p-1}\right)
^{2} \\
\nu _{c,\mathrm{sd}} &=&1.9\times 10^{16}\unit{Hz}L_{0,49}^{13/2}T_{\mathrm{%
sd},3}^{9}M_{\mathrm{ej},-4}^{-8}\epsilon _{B,-1}^{-3/2} \\
F_{\nu ,\max ,\mathrm{sd}} &=&3.5\times 10^{5}\unit{mJy}L_{0,49}^{-1/2}M_{%
\mathrm{ej},-4}^{2}\epsilon _{B,-1}^{1/2}\gamma _{4,4}^{-1}D_{27}^{-2}.
\end{eqnarray}

For Case III the corresponding values are%
\begin{eqnarray}
T_{\mathrm{N1}} &=&2.1\times 10^{-3}\unit{days}L_{0,49}^{-1}M_{\mathrm{ej}%
,-3} \\
T_{\mathrm{dec}} &=&0.9\unit{days}L_{0,49}^{-7/3}T_{\mathrm{sd},3}^{-7/3}M_{%
\mathrm{ej},-3}^{8/3}n^{-1/3} \\
T_{\mathrm{N2}} &=&89.1\unit{days}L_{0,49}^{1/3}T_{sd,3}^{1/3}n^{-1/3} \\
\gamma _{\mathrm{sd}} &=&5.6L_{0,49}T_{\mathrm{sd},3}M_{\mathrm{ej}%
,-3}^{-1}+1,
\end{eqnarray}%
and%
\begin{eqnarray}
\nu _{a,\mathrm{sd}} &=&6.9\times 10^{13}\unit{Hz}L_{0,49}^{-\left(
3p+14\right) /2\left( p+4\right) }T_{\mathrm{sd},3}^{-\left( 3p+14\right)
/\left( p+4\right) }  \notag \\
&&\times M_{\mathrm{ej},-3}^{2\left( p+5\right) /\left( p+4\right) }\epsilon
_{B,-1}^{\left( p+2\right) /2\left( p+4\right) }\gamma _{4,4}^{-2/\left(
p+4\right) } \\
\nu _{m,\mathrm{sd}} &=&2.3\times 10^{15}\unit{Hz}L_{0,49}^{-7/2}T_{\mathrm{%
sd},3}^{-5}M_{\mathrm{ej},-3}^{4}\gamma _{4,4}^{2}\epsilon _{e}^{2}\epsilon
_{B,-1}^{1/2}\left( \frac{p-2}{p-1}\right) ^{2} \\
\nu _{c,\mathrm{sd}} &=&1.8\times 10^{10}\unit{Hz}L_{0,49}^{-3/2}T_{\mathrm{%
sd},3}^{-3}M_{\mathrm{ej},-3}^{2}\epsilon _{B,-1}^{1/2} \\
F_{\nu ,\max ,\mathrm{sd}} &=&4.6\times 10^{3}\unit{Jy}L_{0,49}^{-1/2}T_{%
\mathrm{sd},3}^{-2}M_{\mathrm{ej},-3}^{2}\epsilon _{B,-1}^{1/2}\gamma
_{4,4}^{-1}D_{27}^{-2}.
\end{eqnarray}

\section{Ionization Break Out}

\label{sec:Ion}

\subsection{Opacity Estimate}

Our calculations above are based on the results of \cite{kasen13}, who
studied the opacity at optical/infrared wavelengths, in particular the
bound-bound opacity. For UV radiation, we are interested in the radiation at
early times, i.e. $t\lesssim 600\unit{s}$, because the ejecta will become
transparent at later times. During such early times, the ejecta are hot
enough to thermally ionize the lanthanides, e.g. Ce, to Ce$^{3+}$. For
simplicity, in the following calculations we assume that Ce is completely
thermally ionized as Ce$^{3+}$. As estimated below, the opacity at UV
wavelengths is dominated by the bound-free (photoionization) transitions. We
will therefore calculate the photoionization of Ce$^{3+}$ by the UV
radiation from RS. The forth ionization potential of Ce is $\chi _{4}=36.72%
\unit{eV}$ \citep{Cox01}. We first estimate the different opacities as
follows.

The wavelength independent electron scattering opacity is %
\citep[e.g.,][]{kasen13}%
\begin{equation}
\kappa _{\mathrm{es}}=\frac{\bar{x}\sigma _{T}}{\bar{A}m_{p}}\approx
0.4\left( \frac{\bar{x}}{\bar{A}}\right) \unit{cm}^{2}\unit{g}^{-1},
\label{eq:kappa-es}
\end{equation}%
where $\bar{A}$ is the mean atomic weight of the ions, $\bar{x}$ the mean
ionization fraction, $m_{p}$ the proton mass. After ultraviolet ionization,
Ce is ionized to a level $\bar{x}=4$. With $\bar{A}\approx 140$ we find $%
\kappa _{\mathrm{es}}\approx 0.01\unit{cm}^{2}\unit{g}^{-1}$, which is a
factor of $\sim 20$ smaller than the usual value $\kappa _{\mathrm{es}}=0.2%
\unit{cm}^{2}\unit{g}^{-1}$.

The free-free opacity can be found as \citep[e.g.,][]{Rybicki79,kasen13}
\begin{equation}
\kappa _{\mathrm{ff}}=0.15\frac{\bar{x}^{3}}{\bar{A}^{2}}\rho
_{-8}T_{5}^{-1/2}\lambda _{-5}^{3}\left( 1-e^{-hc/\lambda kT}\right) \unit{cm%
}^{2}\unit{g}^{-1},
\end{equation}%
where the typical density of the ejecta at early times is $\rho \gtrsim
10^{-8}\unit{g}\unit{cm}^{-3}$ and temperature $T\gtrsim 2\times 10^{5}\unit{%
K}\simeq T_{\mathrm{ion}}=\chi _{3}/k$, where $\chi _{3}=20.198\unit{eV}$ is
the third ionization potential of Ce, and $T_{\mathrm{ion}}$ the
corresponding temperature. For the values $\bar{x}$ and $\bar{A}$ mentioned
above, we see that the free-free opacity at UV wavelengths is $\kappa _{%
\mathrm{ff}}\approx 5\times 10^{-5}\unit{cm}^{2}\unit{g}^{-1}$, which is
completely negligible. The free-free opacity at X-ray band is even smaller
because of the dependence of opacity on wavelength $\lambda $.

The bound-bound opacities of the lanthanides are a function of both
temperature and wavelength \citep{kasen13}. On the one hand, the line
expansion opacity of cerium (Ce, $Z=58$, $f$-shell), for example, decreases
to the red and sharply drops to zero for $\lambda \lesssim 1000\unit{%
\text{\AA}%
}$ \citep[Figure 7 in][]{kasen13}. That is, the bound-bound opacity in the
UV and X-ray bands is negligible at the temperature we are interested in,
i.e. $T>T_{\mathrm{ion}}$. On the other hand, the Planck mean expansion
opacity of neodymium (Nd, $Z=60$, $f$-block, similar to Ce in shell
structure) increases sharply with temperature when $T\lesssim 4000\unit{K}$
and then decreases with temperature \citep[Figure 6
in][]{kasen13}. The reason is that, with the increase of temperature, more
excited levels are populated. But if the gas becomes hot enough to ionize,
the originally populated state leaves blank and the opacity declines. The
bound-bound opacity declines sharply when the temperature $T>15000\unit{K}$,
i.e., $\kappa _{\mathrm{bb}}\lesssim 10^{-3}\unit{cm}^{2}\unit{g}^{-1}$ %
\citep[Figure 9 of][]{kasen13}. For the temperature we are interested for UV
radiation, i.e. $T>T_{\mathrm{ion}}$, $\kappa _{\mathrm{bb}}$ becomes
negligibly small.

The bound-free opacity at the threshold energy is given by %
\citep[e.g.,][]{kasen13}
\begin{equation}
\kappa _{\mathrm{bf}}=\frac{\sigma _{0}}{\bar{A}m_{p}}\frac{e^{-\Delta E/kT}%
}{Z\left( T\right) },  \label{eq:kappa-bf}
\end{equation}%
where $Z\left( T\right) $ is the partition function and $\Delta E$ the
excitation energy. The Boltzmann factor $e^{-\Delta E/kT}$ takes account of
the fact that an atom has to be thermally excited so that a photon has
enough energy to ionize the excited atom. The wavelength-dependent parameter
$\sigma _{0}$ is not accurately constrained but can be approximated as the
hydrogenic value $\sigma _{0}\approx 6\times 10^{-18}\unit{cm}^{2}$ at
optical/UV wavelengths. For UV radiation, $\Delta E\approx 0$, unless the
temperature $T\gg T_{\mathrm{ion}}$. The largest uncertainty for the
calculation of $\kappa _{\mathrm{bf}}$ comes from the partition function $%
Z\left( T\right) $. \cite{Irwin81} tabulated $Z\left( T\right) $ for Ce$^{+}$
and Ce$^{2+}$ in the temperature range $1000\unit{K}$-$16000\unit{K}$. The
partition function increases slowly with temperature. It is also expected
that the partition function of Ce$^{3+}$ is smaller than Ce$^{2+}$ at high
temperature. Accordingly, we estimate $Z\left( T\right) \sim 1000$ for $%
T\gtrsim T_{\mathrm{ion}}$. Taking the typical values, we find $\kappa _{%
\mathrm{bf}}\approx 25e^{-\Delta E/kT}\unit{cm}^{2}\unit{g}^{-1}$ for $%
T\gtrsim T_{\mathrm{ion}}$. We therefore conclude that for UV photons,
bound-free opacity is dominant. For X-ray, $\Delta E=0$, but $\sigma _{0}$
takes a value several orders of magnitude smaller than it does at the
optical wavelengths \citep{hakken87}. Consequently, the bound-free opacity
for X-ray is negligible and the opacity at X-ray is dominated by $\kappa _{%
\mathrm{es}}$.

Although the above reasoning is quite robust, but owing to the complex
nature of opacities, it is worth of more scrutiny here. Figure 7 in \cite%
{kasen13} shows that the opacity of osmium (Os, $Z=76$, $d$-shell) is very
high at UV wavelengths. But this high opacity is only for temperature $\sim
5000\unit{K}$. Figure 6 in \cite{kasen13} shows that the opacity of Fe (also
$d$-shell in shell structure, therefore similar to Os) keeps constant in the
temperature range $5000$-$14000\unit{K}$. But with a temperature high enough
to ionize Os, the opacity will decline, as explained above. Consequently we
can tentatively infer that the bound-bound opacity of Os at temperature $T_{%
\mathrm{ion}}$ is smaller than the bound-free opacity evaluated in the
previous paragraph. Another uncertainty comes from actinides ($90<Z<100$).
While the actinide series is generally of very low abundance, but because of
the fact that line expansion opacities are expected to shift to shorter
wavelengths for heavy elements such as actinides, the actinide series may
make significant contribution to the opacity that exceeds the bound-free
opacity estimated above. If the bound-bound opacity contributed by Os and/or
actinides is in excess of the bound-free opacity, we will not observe an UV
ionization break-out.

\subsection{Ionization by ultraviolet radiation from RS}

Before elaborating the calculation of ionization, we need evaluate the
recombination factor, which can be calculated according to the bound-free
opacity as \citep{Rybicki79}

\begin{equation}
\qquad \alpha _{\mathrm{rec}}=4\pi \left( \frac{m_{e}}{2\pi kT}\right) ^{3/2}%
\frac{\sigma _{0}}{m_{e}^{3}c^{2}}\frac{2g_{n}}{g_{e}g_{+}}\frac{\chi ^{3}}{%
Z\left( T\right) }e^{\left( \chi -\Delta E\right) /kT}\func{Ei}\left( \frac{%
\chi }{kT}\right) ,  \label{eq:alpha_rec}
\end{equation}%
where $g_{e}$, $g_{+}$, $g_{n}$ are the statistical weight factors of
electron, Ce$^{4+}$ and Ce$^{3+}$, respectively, $m_{e}$ the electron mass,
and $\func{Ei}\left( x\right) $ is defined as%
\begin{equation}
\func{Ei}\left( x\right) =\int_{x}^{\infty }\frac{e^{-t}}{t}dt.
\end{equation}%
For Ce$^{4+}$ we set $\chi =\chi _{4}$.

The width, $\Delta _{\mathrm{ej},i}^{\prime }$, of the ionized zone (Ce$%
^{4+} $) of the ejecta shell in the comoving frame evolves as
\citep[cf. Equation
(9-8) of][]{Harwit06}%
\begin{equation}
\frac{d\Delta _{\mathrm{ej},i}^{\prime }}{dt^{\prime }}=\left( 4\pi r^{2}\xi
n_{3+}\right) ^{-1}\frac{dN_{i}}{dt^{\prime }}-\Delta _{\mathrm{ej}%
,i}^{\prime }n_{e}\alpha _{\mathrm{rec}},
\end{equation}%
where%
\begin{equation}
n_{3+}=\frac{M_{\mathrm{ej}}}{4\pi r^{2}\xi \Delta _{\mathrm{ej}}^{\prime }%
\bar{A}m_{p}}
\end{equation}%
is the number density of Ce$^{3+}$ before ionization by UV radiation and $%
dN_{i}/dt^{\prime }$ the injection rate of the ionizing photons from RS. The
electron density $n_{e}$ is $n_{e}=\bar{x}n_{4+}$, where $n_{4+}$ $\left(
n_{4+}=n_{3+}\right) $ is the number density of Ce$^{4+}$. By setting $%
n_{4+}=n_{3+}$, we neglect the thermal expansion of the photoionized zone
due to the increase of pressure relative to the Ce$^{3+}$ zone.

The UV photons from RS are scattered off free electrons before ionizing Ce$%
^{3+}$ zone. The optical depth for the ionizing UV photons is%
\begin{equation}
\tau _{\mathrm{UV}}=\bar{A}m_{p}n_{3+}\kappa _{\mathrm{es}}\Delta _{\mathrm{%
ej},i}^{\prime }.
\end{equation}%
The injection rate $dN_{i}/dt^{\prime }$ is therefore evaluated as%
\begin{equation}
\frac{dN_{i}}{dt^{\prime }}=4\pi D_{L}^{2}\frac{F_{\nu ^{\prime }}^{\prime
}e^{-\tau _{\mathrm{UV}}}}{h\nu ^{\prime }}\Delta \nu ^{\prime },
\end{equation}%
where $F_{\nu ^{\prime }}^{\prime }=\mathcal{D}^{-3}F_{\nu }$ is the energy
flux from RS in comoving frame, $h\nu ^{\prime }=\chi _{4}$ is the energy of
ionizing photons. The thermal broadening is given by%
\begin{equation}
\Delta \nu ^{\prime }=\nu ^{\prime }\sqrt{\frac{3kT}{\bar{A}m_{p}c^{2}}}
\end{equation}%
The thickness of the transition from Ce$^{3+}$ zone to Ce$^{4+}$ zone is of
the order of mean free ionizing path $\delta =\left( n_{3+}\sigma _{\mathrm{%
bf}}\right) ^{-1}\simeq 1.5\times 10^{6}\unit{cm}n_{3+,14}^{-1}$, which is
more than one order of magnitude smaller than the width of the ejecta $%
\Delta _{\mathrm{ej}}^{\prime }\gtrsim 10^{7}\unit{cm}$.

The calculated UV light curves and ionization fraction ($\Delta _{\mathrm{ej}%
,i}^{\prime }/\Delta _{\mathrm{ej}}^{\prime }$) are depicted in Figure \ref%
{fig:ionization}. The observed UV flux is attenuated by a factor $e^{-\tau _{%
\mathrm{bf}}}$, where $\tau _{\mathrm{bf}}=\kappa _{\mathrm{bf}}\left(
\Delta _{\mathrm{ej}}^{\prime }-\Delta _{\mathrm{ej},i}^{\prime }\right)
\bar{A}m_{p}n_{3+}$. Before the ejecta are completely photoionized, $\tau _{%
\mathrm{bf}}\gtrsim 100$ so that effectively the UV flux can be observed
only when the ejecta are completely photoionized, i.e. ionization break out.
This is why we see an abrupt increase of UV flux when the ionization
fraction equals 1. As the ejecta expand, the ejecta temperature drops
rapidly so that when $T=T_{\mathrm{ion}}$ the thermally ionized Ce$^{3+}$
begins to recombine. We see from Figure \ref{fig:ionization} that it is not
long when the ejecta become transparent after the temperature drops below $%
T_{\mathrm{ion}}$. As a result, our treatment of the ultraviolet flux is not
affected very much by the recombination.

\section{Discussion}

\label{sec:discuss}

$R$-process material is of great importance astrophysically, the studies of
which are unfortunately very difficult in laboratory. It is therefore
particularly valuable if we can study their properties by observing
radiation, obscured by a pure $r$-process material, at different
wavelengths. To this aim, we propose in this paper to observe the RS
emission obscured by the ejecta, which is believed to be a pure $r$-process
material.

Figures \ref{fig:Case-I}-\ref{fig:Case-III} show that the ejecta shell is
subrelativistic when it becomes transparent (at time $t_{\mathrm{x,thin}}$)
to X-ray emission from RS. During this phase the radius of the ejecta shell,
suitable for Case I, II, and III, is%
\begin{equation}
r=2.1\times 10^{13}\unit{cm}L_{0,47}^{1/2}M_{\mathrm{ej}%
,-4}^{-1/2}t_{3}^{3/2},\qquad t<T_{N1}
\end{equation}%
from which $t_{\mathrm{x,thin}}$ can be estimated as%
\begin{equation}
t_{\mathrm{x,thin}}=7.1\times 10^{2}\unit{s}M_{\mathrm{ej}%
,-4}^{2/3}L_{0,47}^{-1/3},
\end{equation}%
upon substituting Equation $\left( \ref{eq:kappa-es}\right) $. Inspection of
Figures \ref{fig:Case-I}-\ref{fig:Case-III} indicates that $t_{\mathrm{x,thin%
}}$\ is approximately the time when the X-ray flux peeks from the left edge
of the light curve. This behavior is also true for optical flux. Note that
this estimate for $t_{\mathrm{x,thin}}$ is accurate enough in comparison
with the numerically determined value.

The estimate for the time, $t_{\mathrm{opt,thin}}$, when the ejecta become
optically thin for visible light can be done in a similar way as long as we
are aware of the fact that at this time the ejecta shell is in relativistic
regime so that the radius should be given by%
\begin{equation}
r=1.3\times 10^{13}\unit{cm}L_{0,47}^{2}M_{\mathrm{ej},-4}^{-2}t_{3}^{2},%
\qquad T_{N1}<t<\min \left( T_{\mathrm{sd},}T_{\mathrm{dec}}\right)
\end{equation}%
from which we get%
\begin{equation}
t_{\mathrm{opt,thin}}=2.2\times 10^{3}\unit{s}\kappa _{\mathrm{bb}}^{1/6}M_{%
\mathrm{ej},-4}^{5/6}L_{0,47}^{-2/3}.
\end{equation}%
Unfortunately, $t_{\mathrm{opt,thin}}$ depends on $\kappa _{\mathrm{bb}}$
very weakly so that only a loose constraint on $\kappa _{\mathrm{bb}}$ can
be obtained. If, on the other hand, the ejecta are massive enough and/or the
spin-down luminosity of the central magnetar is moderate so that the ejecta
shell is still in the subrelativistic regime when it becomes transparent for
visible light, we have $t_{\mathrm{opt,thin}}\propto \kappa _{\mathrm{bb}%
}^{1/3}$ and $\kappa _{\mathrm{bb}}$ can be constrained more compactly.

To evaluate the dependence of $t_{\mathrm{UV,BO}}$, the break-out time of UV
radiation, on $\kappa _{\mathrm{bf}}$, we vary $\kappa _{\mathrm{bf}}$ by a
factor $f$, i.e., $\kappa _{\mathrm{bf}}^{\prime }=f\kappa _{\mathrm{bf}}$.
From Equations $\left( \ref{eq:kappa-bf}\right) $ and $\left( \ref%
{eq:alpha_rec}\right) $ we see that the factor $f$ contains the
uncertainties in $\sigma _{0}$ and $Z\left( T\right) $. The uncertainty in $%
\sigma _{0}$\ is a few, so is $Z\left( T\right) $. So we expect the
uncertainty of $f$\ is $\sim 10$. The solid lines from left to right in the
insets of Figure \ref{fig:ionization} show the UV light curves with $f=0.1$,
$1$, and $10$, respectively. This time lies between $t_{\mathrm{x,thin}}$
and $t_{\mathrm{opt,thin}}$, as it should be.

This model contains many free parameters to be constrained by fitting light
curves. \cite{wang13b} demonstrated that by simultaneously fitting the
optical and radio light curves, one can accurately constrain the spin-down
luminosity $L_{\mathrm{sd}}$, the spin-down time scale $T_{\mathrm{sd}}$,
the ejecta mass $M_{\mathrm{ej}}$, the density of ambient medium (region 1) $%
n$, the Lorentz factor of leptonized wind (region 4) $\gamma _{4}$, the
electron power-law index of reverse shocked wind (region 3) $p^{\mathrm{RS}}$%
, the magnetic energy fraction $\epsilon _{B}$\ and the redshift $z$\ of the
source \citep[Table 2 of][]{wang13b}.

\cite{wang13b} also determined the launch time of the Poynting flux of
PTF11agg. This is possible because the launch time affects the temporal
decline index of the early optical light curve. To accurately determine the
launch time of the Poynting flux by timing X-ray flux, Figures \ref%
{fig:Case-I}-\ref{fig:Case-III} indicate that we have to take observation in
X-ray band following the merger no later than $\sim 100$-$700\unit{s}$. A
similar observational strategy has to be employed to determine the
ionization break-out of ultraviolet light (see Figure \ref{fig:ionization}).
Future gravitational observation of NSMs may also help determine the launch
time of Poynting flux because this time roughly coincides with the merger
time. Figure \ref{fig:ionization} shows that the ionization break-out time
depends on $\kappa _{\mathrm{bf}}$\ weakly. Thus by timing X-ray, UV, and
visual light we can obtain weak constraints on $\kappa _{\mathrm{bb}}$\ and $%
\kappa _{\mathrm{bf}}$.

\section{Conclusions}

\label{sec:conclusion}

\begin{figure}[tbph]
\centering\includegraphics[width=1\textwidth,angle=0]{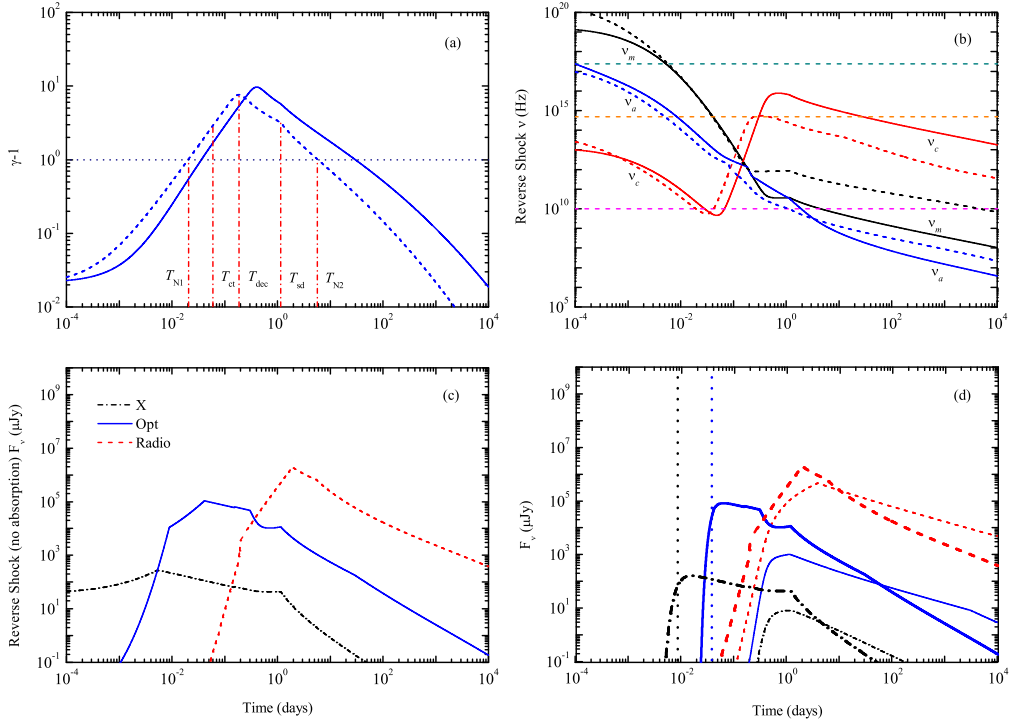}
\caption{Calculation results for $L_{\mathrm{sd,}i}=10^{47}\unit{erg}\unit{s}%
^{-1}$, $T_{\mathrm{sd}}=10^{5}\unit{s}$ and $M_{\mathrm{ej}%
}=10^{-4}M_{\odot }$. (a) The evolution of Lorentz factor. (b) The
characteristic frequencies of RS emission. The three dashed lines mark the
X-ray, optical ($R$) and radio (8 GHz) bands, respectively. (c) Light curves
of RS without absorption by ejecta. (d) Light curves of RS with absorption
by ejecta considered (thick) and of FS (thin). The two vertical dotted lines
in panel (d) mark the times when the ejecta become transparent for X-ray and
visual light respectively. The emission by the heated ejecta is negligible
in this case, so we do not show them. Dashed lines in panels (a) and (b) are
the results based on the dynamics expressed by Equation $\left( \protect\ref%
{eq:PTF11agg-dyn}\right) $. In panel (a) $T_{\mathrm{N1}}$ and $T_{\mathrm{N2%
}}$ are the times when $\protect\gamma -1=1$, i.e., the transition time
between relativistic motion and Newtonian dynamics, $T_{\mathrm{ct}}$ is the
transition time for cooling Lorentz factor, see \protect\cite{wang13b} for
more explanation. In the calculations we set initial ejecta velocity $%
\protect\beta _{\mathrm{ej},0}=0.2$.}
\label{fig:Case-I}
\end{figure}

\begin{figure}[tbph]
\centering\includegraphics[width=1\textwidth,angle=0]{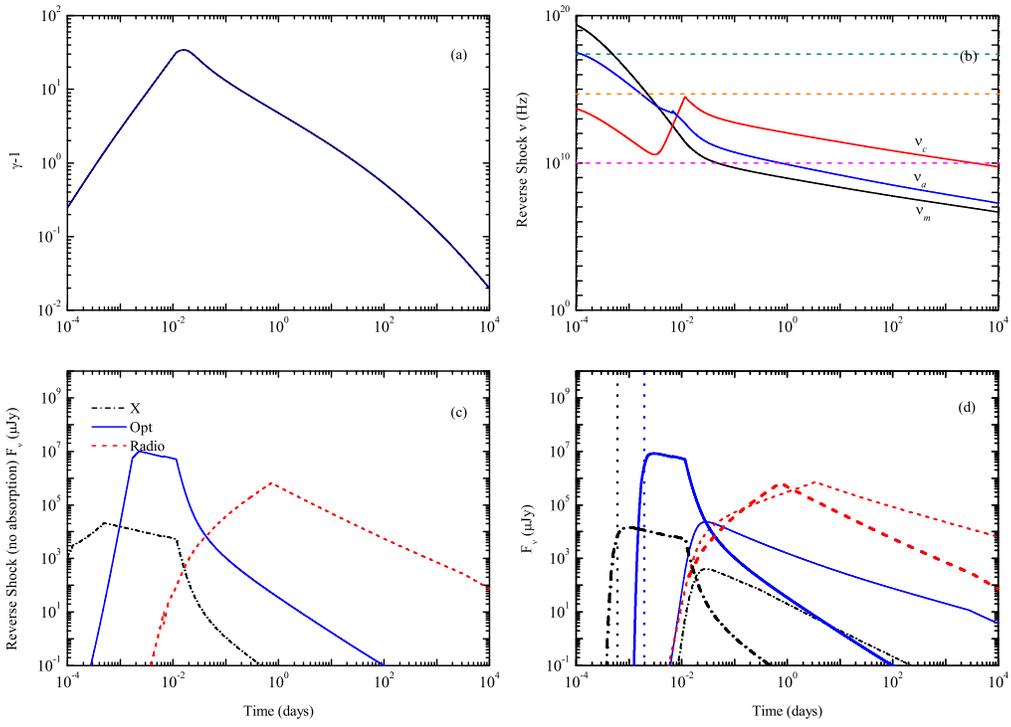}
\caption{Calculation results for $L_{\mathrm{sd},i}=10^{49}\unit{erg}\unit{s}%
^{-1}$, $T_{\mathrm{sd}}=10^{3}\unit{s}$, $M_{\mathrm{ej}}=10^{-4}M_{\odot }$%
. The meanings of the panels are the same as in Figure \protect\ref%
{fig:Case-I}. Ejecta emission is negligible.}
\label{fig:Case-II}
\end{figure}

\begin{figure}[tbph]
\centering\includegraphics[width=1\textwidth,angle=0]{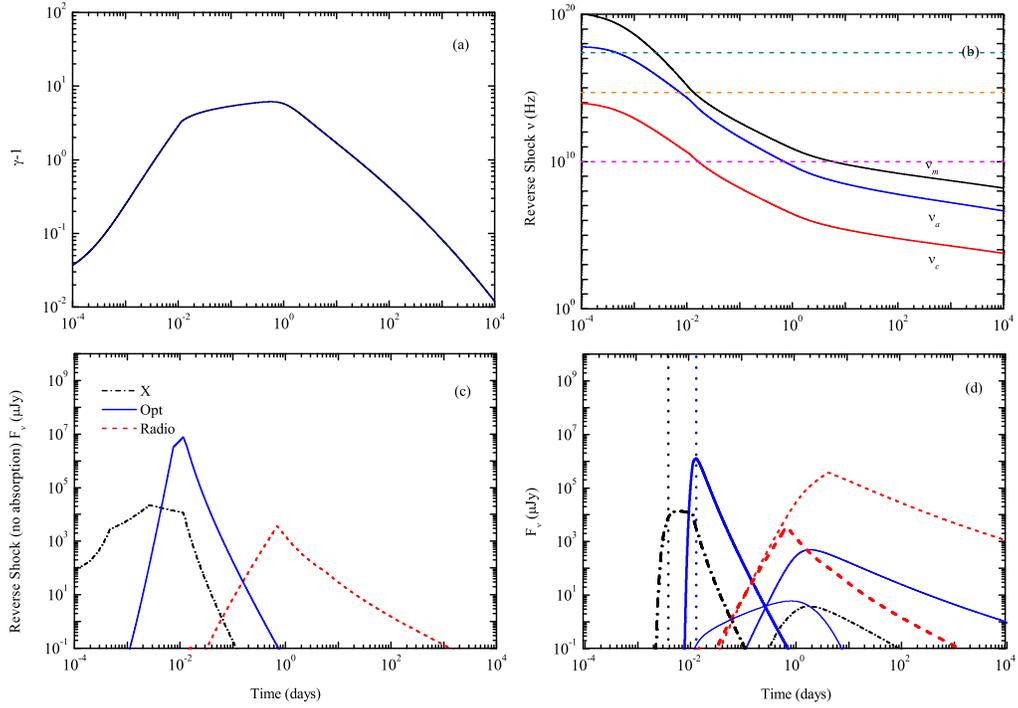}
\caption{Calculation results for $L_{\mathrm{sd},i}=10^{49}\unit{erg}\unit{s}%
^{-1}$, $T_{\mathrm{sd}}=10^{3}\unit{s}$, $M_{\mathrm{ej}}=10^{-3}M_{\odot }$%
. The meanings of the panels are the same as in Figure \protect\ref%
{fig:Case-I}. In panel (d) the optical emission by the ejecta is shown as
thin line. Ejecta emission in other bands is negligible. This is the only
case discussed in this paper that the ejecta emission becomes appreciable in
optical band.}
\label{fig:Case-III}
\end{figure}

\begin{figure}[tbph]
\centering%
\subfigure[$M_{\mathrm{ej}}=10^{-4}M_{\odot },L_{\mathrm{sd},i}=10^{47}\unit{erg}\unit{s}^{-1}$]{
\includegraphics[width=0.48\textwidth,angle=0]{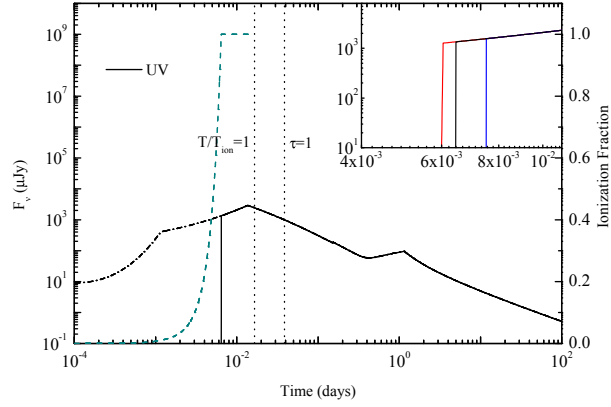}} \hspace{0.3cm}
\subfigure[$M_{\mathrm{ej}}=10^{-4}M_{\odot },L_{\mathrm{sd},i}=10^{49}\unit{erg}\unit{s}^{-1}$]{
\includegraphics[width=0.48\textwidth,angle=0]{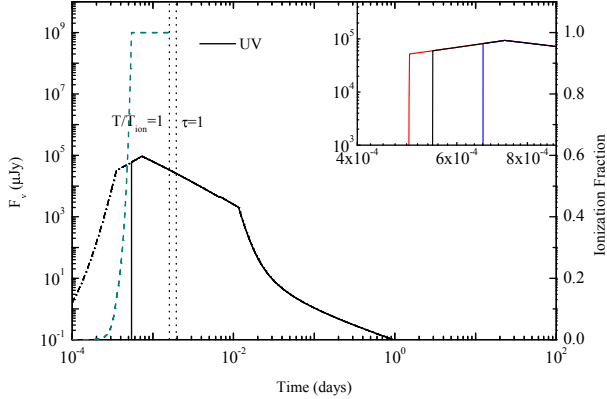}}%
\subfigure[$M_{\mathrm{ej}}=10^{-3}M_{\odot },L_{\mathrm{sd},i}=10^{49}\unit{erg}\unit{s}^{-1}$]{
\includegraphics[width=0.48\textwidth,angle=0]{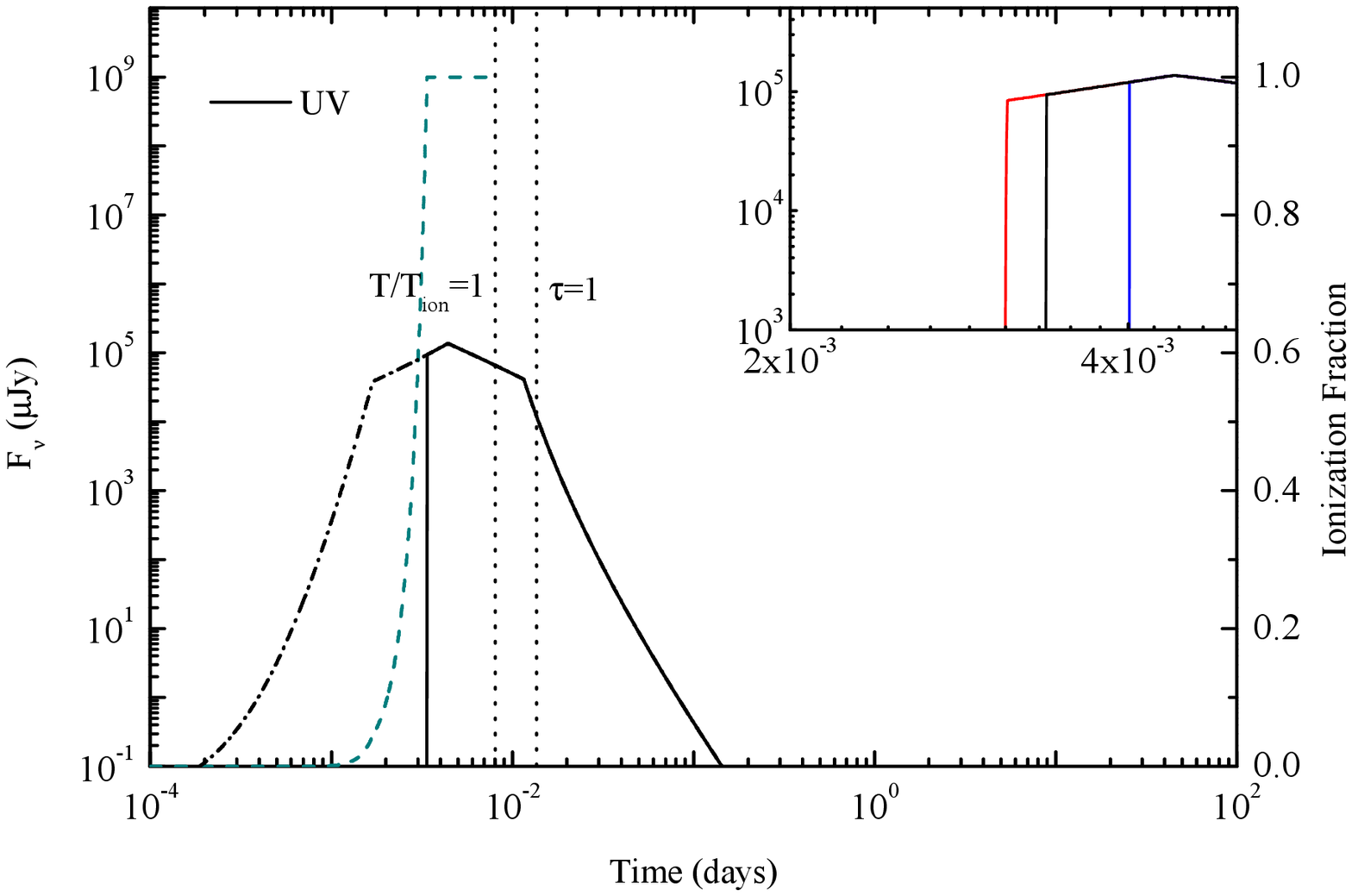}}
\caption{{}Ultraviolet radiation from RS. The solid lines correspond to the
observed ultraviolet flux leaked from the ionizing ultraviolet flux from RS
(dash-dotted lines). The dashed lines are the ionization fraction of the
ejecta. The vertical dotted lines indicate the times when $T/T_{\mathrm{ion}%
}=1$ and $\protect\tau =1$. The solid lines from left to right in the insets
are the UV light curves by taking the bound-free opacity uncertainty factor $%
f=0.1$, $1$, and $10$, respectively.}
\label{fig:ionization}
\end{figure}

In this paper, we have analyzed electromagnetic counterparts in various
bands to the binary neutron star merger, and arrived at the following
conclusions:

\begin{enumerate}
\item[(i)] By assuming that the Poynting flux of the central magnetar
becomes lepton-dominated so that RS develops, broad-band EM signals, i.e.
radio, optical, UV, and X-ray emission, can be produced. The NSM ejecta can
be accelerated to relativistic speed so that a forward shock can form. We
therefore expect to observe EM signals first from RS and subsequently from
FS.

\item[(ii)] In the study of the optical and radio radiation of PTF11agg, the
simple dynamics, i.e., Equation $\left( \ref{eq:PTF11agg-dyn}\right) $, was
adopted by \cite{wang13b}. In this paper, the dynamics is determined by
pressure balance and it is found that Equation $\left( \ref{eq:PTF11agg-dyn}%
\right) $ is an acceptable approximation.

\item[(iii)] In studying PTF11agg, \cite{wang13b} ignored the absorption of
optical radiation by the ejecta. We find in this paper, by including the
absorption effect, that by the time the first optical datum was taken, the
ejecta become optically thin so that the treatment of \cite{wang13b} is
justified. \cite{wang13b} also ignored the emission from the ejecta itself,
which is verified in this paper (Figure \ref{fig:Case-I}) that the emission
of the ejecta itself is negligible in all relevant bands.

\item[(iv)] Scattering off free electrons will obscure the early X-ray
emission from RS. Optical emission from RS will be suppressed by bound-bound
opacity until the ejecta become transparent at later times. UV radiation
from RS, on the other hand, has the right energy to ionize the hot ejecta
during its early expansion and ionization breakout should therefore be
observed.

\item[(v)] Given the weak dependence of arrival times of EM signals at
different wavelengths on different opacities, we conclude that the timing of
X-ray, UV, and optical light is not an accurate method to observationally
constrain the opacities of $r$-process material.

\item[(vi)] Recent papers by \cite{gao13} and \cite{wang13b} make an
approximation that the merger ejecta is rapidly compressed into a thin
expanding shell, similar to the one which forms at GRB afterglows. In this
paper we estimate the time it takes for the flux to compress the ejecta to a
thin shell and find that the time is negligible compared with the activity
duration of the central magnetar. As a result, the models proposed by \cite%
{gao13} and \cite{wang13b} can be applied safely.
\end{enumerate}

\begin{acknowledgements}
We are grateful to the referee for insightful comments and constructive suggestions,
 which significantly improve the presentation of this paper. This work is supported by
 the National Basic Research Program (``973" Program)
 of China under Grant No. 2014CB845800. L.J.W. and Z.G.D are also supported by the
 National Natural Science Foundation of China (grant No. 11033002), and Y.W.Y.
 by grant No. 11473008.
\end{acknowledgements}

\clearpage
\begin{table}[tbp]
\caption{Analytical Temporal Scaling Indices of Various Parameters of the
RS. The Analytical Results for Case I were Presented in \protect\cite%
{wang13b}.}
\label{tbl:indices}
\begin{center}
$%
\begin{tabular}{lcccccc}
\hline\hline
& $\gamma -1$ & $r$ & $\nu _{a}$ & $\nu _{m}$ & $\nu _{c}$ & $F_{\nu ,\max }$
\\ \hline
\multicolumn{7}{c}{Case II: $L_{0}\sim 10^{49}\unit{erg}\unit{s}^{-1},M_{%
\mathrm{ej}}\sim 10^{-4}M_{\odot }$} \\ \hline
$t<T_{N1}$ & $1$ & $\frac{3}{2}$ & $-\frac{3p+14}{2\left( p+4\right) }$ & $-%
\frac{3}{2}$ & $-\frac{3}{2}$ & $-\frac{1}{2}$ \\
$T_{N1}<t<T_{\mathrm{ct}}$ & $1$ & $3$ & $-\frac{3p+14}{p+4}$ & $-5$ & $-3$
& $-2$ \\
$T_{\mathrm{ct}}<t<T_{ac}$ & $1$ & $3$ & $-\frac{3p+2}{p+4}$ & $-5$ & $9$ & $%
-2$ \\
$T_{ac}<t<T_{\mathrm{sd}}$ & $1$ & $3$ & $-\frac{5p+12}{p+4}$ & $-5$ & $9$ &
$-2$ \\
$T_{\mathrm{sd}}<t<T_{N2}$ & $-\frac{3}{8}$ & $\frac{1}{4}$ & $-\frac{9p+46}{%
16\left( p+4\right) }$ & $-\frac{9}{16}$ & $-\frac{17}{16}$ & $-\frac{9}{16}$
\\
$T_{N2}<t<T_{am2}$ & $-\frac{6}{5}$ & $\frac{2}{5}$ & $-\frac{3p+14}{5\left(
p+4\right) }$ & $-\frac{3}{5}$ & $-\frac{3}{5}$ & $-\frac{3}{5}$ \\
$T_{am2}<t$ & $-\frac{6}{5}$ & $\frac{2}{5}$ & $-\frac{18}{25}$ & $-\frac{3}{%
5}$ & $-\frac{3}{5}$ & $-\frac{3}{5}$ \\ \hline
\multicolumn{7}{c}{Case III: $L_{0}\sim 10^{49}\unit{erg}\unit{s}^{-1},M_{%
\mathrm{ej}}\sim 10^{-3}M_{\odot }$} \\ \hline
$t<T_{N1}$ & $1$ & $\frac{3}{2}$ & $-\frac{3p+14}{2\left( p+4\right) }$ & $-%
\frac{3}{2}$ & $-\frac{3}{2}$ & $-\frac{1}{2}$ \\
$T_{N1}<t<T_{\mathrm{sd}}$ & $1$ & $3$ & $-\frac{3p+14}{p+4}$ & $-5$ & $-3$
& $-2$ \\
$T_{\mathrm{sd}}<t<T_{\mathrm{dec}}$ & $0$ & $1$ & $-\frac{3p+14}{2\left(
p+4\right) }$ & $-\frac{3}{2}$ & $-\frac{3}{2}$ & $-\frac{3}{2}$ \\
$T_{\mathrm{dec}}<t<T_{N2}$ & $-\frac{3}{8}$ & $\frac{1}{4}$ & $-\frac{9p+46%
}{16\left( p+4\right) }$ & $-\frac{9}{16}$ & $-\frac{9}{16}$ & $-\frac{9}{16}
$ \\
$T_{N2}<t$ & $-\frac{6}{5}$ & $\frac{2}{5}$ & $-\frac{3p+14}{5\left(
p+4\right) }$ & $-\frac{3}{5}$ & $-\frac{3}{5}$ & $-\frac{3}{5}$ \\ \hline
\end{tabular}%
$%
\end{center}
\end{table}

\end{document}